**Spontaneous Exciton Dissociation in Transition Metal Dichalcogenide Monolayers**

Taketo Handa[1], Madisen A. Holbrook[2], Nicholas Olsen[1], Luke N. Holtzman[3], Lucas Huber[1], Hai I. Wang[4], Mischa Bonn[4], Katayun Barmak[3], James C. Hone[5], Abhay N. Pasupathy[2], X.-Y. Zhu[1,†]

[1] Department of Chemistry, Columbia University, New York, New York 10027, USA.

[2] Department of Physics, Columbia University, New York, New York 10027, USA.

[3] Department of Applied Physics and Applied Mathematics, Columbia University, New York, NY 10027, USA

[4] Max Planck Institute for Polymer Research, 55128 Mainz, Germany

[5] Department of Mechanical Engineering, Columbia University, New York, NY 10027, USA

[†]To whom correspondence should be addressed. E-mail: xyzhu@columbia.edu.

**Since the seminal work on $MoS_2$ monolayers[1,2], photoexcitation in atomically-thin transition metal dichalcogenides (TMDCs) has been assumed to result in excitons with large binding energies (~ 200-600 meV) [3,4]. Because the exciton binding energies are order-of-magnitude larger than thermal energy at room temperature, it is puzzling that photocurrent and photovoltage generation have been observed in TMDC-based devices[5–9], even in monolayers with applied electric fields far below the threshold for exciton dissociation[10]. Here, we show that the photoexcitation of TMDC monolayers results in a substantial population of free charges. Performing ultrafast terahertz (THz) spectroscopy [11,12] on large-area, single crystal $WS_2$, $WSe_2$, and $MoSe_2$ monolayers[13], we find that ~10% of excitons spontaneously dissociate into charge carriers with lifetimes exceeding 0.2 ns. Scanning tunnelling microscopy reveals that photo-carrier generation is intimately related to mid-gap defect states, likely via trap-mediated Auger scattering. Only in state-of-the-art quality monolayers[14], with mid-gap trap densities as low as $10^9$ $cm^{-2}$, does intrinsic exciton physics start to dominate the THz response. Our findings reveal that excitons or excitonic complexes are only the predominant quasiparticles in photo-excited TMDC monolayers at the limit of sufficiently low defect densities.**

**Main text**

Excitons in atomically-thin two-dimensional (2D) semiconductors, particularly TMDCs, are central to a broad range of problems in 2D materials research. Examples include light-matter interactions[15,16], optoelectronic processes[17,18], photocatalysis[19], quantum phases[20,21], and sensors for quantum phenomena[22–24]. In all these examples, photophysical properties of TMDC monolayers are ultimately dictated by whether optical excitation primarily creates excitons[3,4,17] or free charge carriers[5,7,8]. Terahertz time-domain spectroscopy (THz-TDS) is a powerful technique that can investigate the nature of the photoexcited species by providing a contact-free measurement of charge carrier properties, owing to the distinct THz spectroscopic signatures of excitons and free charges[11,12]. Previous studies have applied THz-TDS to investigate excited-state properties in monolayer TMDCs, but the reported amplitudes, timescales, and even the sign of THz signal vary greatly from report to report[25–29]. One complication is that many of these experiments employed excitation densities beyond the Mott phase transition, causing excitons to dissociate into electron/hole plasmas[30–32]. Moreover, the long wavelength of THz radiation necessitates the use of large area samples that could previously only be obtained from chemical vapor deposition, which typically yields polycrystalline monolayers with reduced crystalline quality compared to mechanically exfoliated single crystal monolayers. The preparation of single-crystal TMDC monolayers is crucial when investigating their photophysics because, as we show in the present study, the fate of excitons in TMDC monolayers depends critically on defect density.

We use a gold-tape exfoliation technique[13] to prepare single-crystal TMDC monolayers with macroscopic areas (mm-cm). Fig. 1a-i shows optical image of a representative single crystal $MoSe_2$ monolayer; here the photo is obtained on a Si wafer with 285 nm $SiO_2$ for enhanced optical contrast. In all spectroscopic measurements, the monolayers are on quartz substrates (see Supplementary Fig. 1 for optical images). The large sizes of the single crystal monolayers, coupled with a high-sensitivity detection scheme[33], allow us to carry out optical pump – THz probe experiments at excitation densities as low as two-orders of magnitude below the Mott threshold. We compare TMDC monolayers ($WS_2$, $WSe_2$, $MoSe_2$) exfoliated from single crystals grown by chemical vapor transport (CVT) and the self-flux method. The latter is known to yield single crystals with defect densities order(s)-of-magnitude lower than those from the former[14]. In this experiment, we photoexcite a macroscopic TMDC monolayer and measure the change in optical conductivity with a time-delayed ($\Delta t$) THz pulse by recording the real-time waveform $E_{THz}$ via electro-optic sampling (EOS) as a function of timing $\tau_{EOS}$ of the electric field, Fig. 1a-ii. Fourier transform of the $\tau_{EOS}$-dependent $E_{THz}$ gives a frequency range of ~0.5-3.5 THz (Fig. 1a-iii). We excite the TMDC monolayers at photon energy either in resonance or above the 1s A-exciton transition, as shown by the arrows on the optical absorption spectra obtained for the macroscopic monolayer samples, in Fig. 1b. We keep the excitation densities below the Mott density of $\sim10^{13}$ $cm^{-2}$ for TMDC monolayers[31,32]. The THz field predominantly detects charge carriers but not charge-neutral excitons as we discuss below.

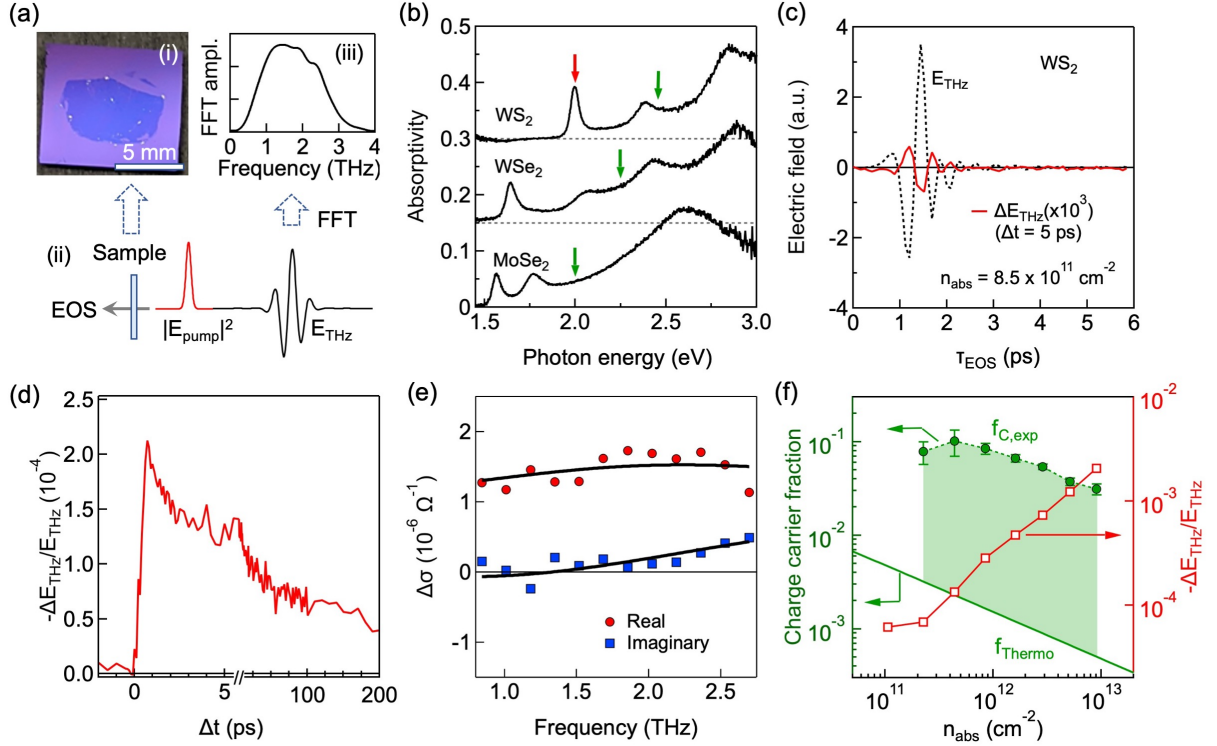

**Fig. 1 | Observation of charge carriers in photoexcited monolayer WS₂.** (a-i) Optical image of representative macroscopic MoSe₂ monolayer on SiO₂/Si substrate for enhanced optical contrast. (a-ii) Schematic of experimental setup. Optical pump excites a monolayer TMD on z-cut quartz, while a time-delayed (Δt) THz probe pulse transmits through the sample and is directed to the electro-optical sampling (EOS) system. The real-time THz waveform $E_{THz}(\tau_{EOS})$ is measured by scanning another gate pulse with the timing $\tau_{EOS}$. (a-iii) Frequency-domain spectrum THz probe obtained via Fourier transform. (b) Absorption spectra of TMDC monolayers. The arrows indicate excitation photon energies. (c) THz transient (dashed) and its photoinduced change (solid) for CVT WS₂ monolayer at absorbed photon density of $8.5 \times 10^{11}$ cm⁻² under above-gap excitation. (d) The photoconductivity dynamics monitored at the fixed $\tau_{EOS}$, corresponding to the maximum of $-E_{THz}/E_{THz}$ under above-gap excitation. (e) Complex photoconductivity spectra together with the fitting result with the Drude-Smith model (see text). (f) Excitation-density dependence of the maximum THz photoconductivity amplitude (red squares, right axis) and corresponding charge-carrier fraction (green circles, left axis). The solid line is the carrier fraction estimated for a perfect monolayer with the thermodynamic Saha equation at 293 K. All experiments carried out at sample temperature of 293 K, unless otherwise noted.

Distinct photoconductivity below the Mott threshold in $WS_2$ monolayers

We first show that in monolayer $WS_2$ exfoliated from a CVT-grown crystal, THz photoconductivity reveals the generation of charge carriers from photo-excitation at photon energies above (green arrow in Fig. 1b) and in resonance with (red arrow in Fig. 1b) the 1s-A exciton transition[34]. Fig. 1c shows the THz electric field trace, $E_{THz}$, transmitted through the $WS_2$ monolayer (black dotted curve) and its pump-induced change, $\Delta E_{THz}$ (red solid curve). The latter is obtained at absorbed photon density $n_{abs} = 8.5 \times 10^{11}$ $cm^{-2}$ and at pump-probe delay $\Delta t = 5$ ps, with excitation photon energy $h\nu = 2.40$ eV (above-gap). The photoinduced transient $\Delta E_{THz}$ tracks the original waveform of $E_{THz}$, but with flipped sign and with no detectable phase shift, a signature of the spectrally broad absorptive response[12]. The photo-induced THz field change at peak $\Delta E_{THz}$ rises on ultrashort time scales ($\Delta t \leq 1ps$) following photoexcitation and extends beyond 200 ps, Fig. 1d, with decay time constants in the 20 – 260 ps range (obtained with bi-exponential fit to data at $\Delta t \geq 10$ ps). The photoinduced complex conductivity spectrum $\Delta\sigma$ (Fig. 1e) obtained from $\Delta E_{THz}$ (see Methods) is dominated by the real part, over the entire $\Delta t$ range (see Supplementary Fig. S2)[12]. The large exciton binding energy and associated high frequency for intra-exciton Rydberg transitions[35], make that excitons do not contribute to the real part of $\Delta\sigma$ in this frequency range (see Supplementary Note 1 and Fig. S3). As such, the absorption of THz must be free-carrier-induced.

We quantify the absolute carrier density through analysis of the complex $\Delta\sigma$ by fitting it to the Drude-Smith model[36] at an effective carrier mass of $m_{eff} \sim 0.35m_e$ ($m_e$ is the bare electron mass, see Methods)[37]. The fitting (solid curves in Fig. 1e) gives a carrier density of $n_c = 7.2\pm1.0 \times 10^{10}$ $cm^{-2}$ together with the Smith parameter of -0.58±0.05 and scattering time of 49±7 fs. Accordingly, the fraction of free carriers to the absorbed photon density $n_{abs}$ is calculated to be $f_c$ ($= n_c/n_{abs}$) = 0.085±0.011. As we discuss below, the generation of free carriers likely originates from the trapping of one carrier and the release of a free carrier from an exciton. Thus, $f_c$ is also the fraction of dissociated excitons. Note that the $f_c$ value obtained at $\Delta t = 5$ ps provides a lower limit estimate since ultrafast recombination may occur in this early time window[38]. The THz conductivity is observed in a broad range of excitation densities down to $10^{11}$ cm$^{-2}$, red open squares (right axis) in Fig. 1f (also see Supplementary Fig. S4 for the complex conductivity spectra). We obtain $f_c$ as a function of n$_{abs}$ (filled green circles, left axis in Fig. 1e). Also shown as solid green line is the calculated fraction of exciton dissociation for monolayer WS$_2$ at room temperature from thermodynamic considerations (see Methods and Supplementary Note 2)[11]. The measured charge carrier fraction from THz measurement is two orders-of-magnitude larger (green shaded region) than the thermodynamic prediction for a perfect monolayer.

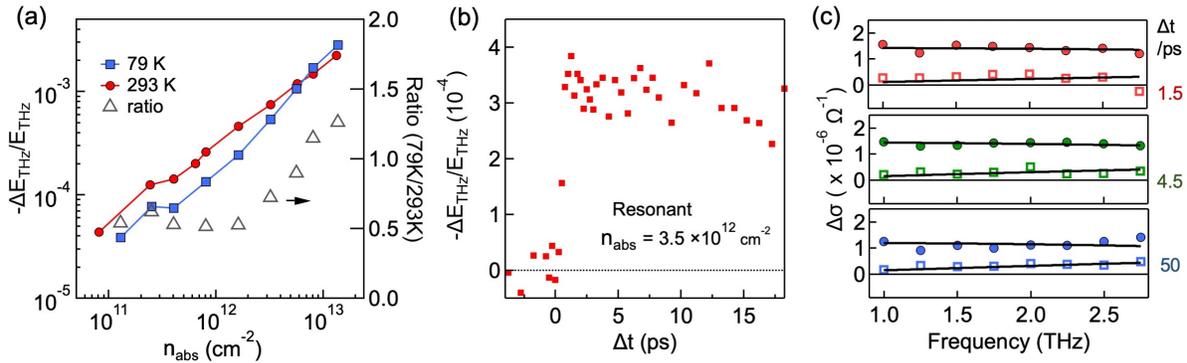

**Fig. 2 | Non-thermal origin for exciton dissociation in CVT WS$_2$.** (a) Excitation density dependence of THz photoconductivity amplitude at 293 and 79 K under the above-gap excitation, showing almost temperature-independent conductivity response. (b) THz conductivity dynamics following the 1s-resonant excitation. (c) Complex conductivity spectra at different pump probe delays, demonstrating that at all delays real part dominates, testifying to the presence of free charges.

We confirm the non-thermal origin of the observed THz photoconductivity by measuring -$\Delta$E$_{THz}$/E$_{THz}$ (fixed $\tau_{EOS}$) at two temperatures, 293 and 79 K, with above-gap excitation conditions in the CVT WS$_2$ monolayer. The amplitudes of photoconductivity and their corresponding $f_c$ values (~0.1) are similar at these two temperatures (see Supplementary Fig. S5). If one considers the thermodynamics of exciton ionization at T = 79 K[11], the charge carrier fraction is predicted to be only $10^{-10}$ to $10^{-11}$ at the relevant excitation densities (Supplementary Fig. S6). Moreover, we observe clear photoconductivity under resonant excitation of the 1s-exciton, Fig 2b. The prompt rise in photoconductivity indicates spontaneous exciton dissociation without the need for excess energy. The real part again dominates the conductivity at different pump-probe delays (Fig. 2c). These results establish that the driving force for exciton dissociation into free charges is not thermal, but likely mediated by an extrinsic process.

<u>Exciton dissociation and mid-gap states in MoSe₂ and WSe₂ monolayers</u>

To understand the origins of carrier generation in photo-excited TMDC monolayers, we investigate the potential role of mid-gap trap states in meeting the energetic requirements for exciton dissociation [39,40]. Specifically, we compare MoSe₂ and WSe₂ monolayers exfoliated from crystals grown by CVT and flux methods; defect densities from the latter are known to be order(s)-of-magnitude lower than those from the former[14,41]. We first focus on MoSe₂ and quantify defect density and its electronic structure using scanning tunneling microscopy and spectroscopy (STM and STS). Fig. 3a shows an STM topographic image of CVT MoSe₂, which features bright defects at a density of $n_d = 2.7 \pm 1.2 \times 10^{10}$ cm⁻²; a magnified image of one such defect is shown in Fig. 3b. For comparison, the density of bright defects in flux-grown MoSe₂ is an order of magnitude lower, at $n_d = 3.6 \pm 2.5 \times 10^9$ cm$^{-2}$. While the exact origin of these bright defects is not known, STS reveals the presence of mid-gap state on each bright defect feature (red curve in Fig. 3c). For comparison, STS shows no mid-gap state (blue curve in Fig. 3c) when the STM tip is not located on the bright defect of CVT MoSe$_2$. As such, the flux MoSe$_2$ has less mid-gap defects and is largely trap-free as exemplified by a representative STS (black curve) in Fig 3c.

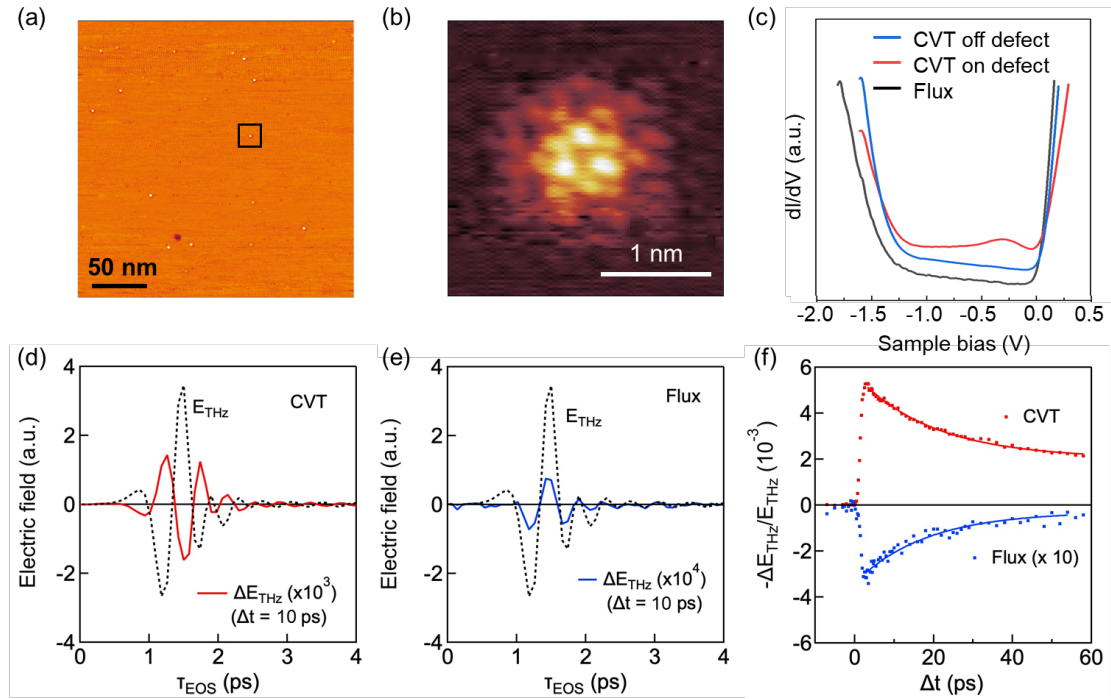

**Fig. 3 | Mid-gap state induced exciton dissociation in CVT MoSe₂ and trion-induced negative photoconductivity in flux MoSe₂.** (a) STM topographic image for CVT MoSe₂, showing bright defects. (b) A close-up image of the bright defect in panel (a). STM images in (a) and (b) were obtained at a sample bias of -1.3 V and -1.5 V, respectively. (c) STS spectra collected on (red) and off (blue) the bright defect in CVT MoSe₂, identifying the existence of mid-gap states on the bright defect. Also shown with black curve is the STS scan for flux grown MoSe₂. (d,e) Photoinduced THz transients (solid lines) for CVT and flux MoSe₂ monolayers, respectively, at Δt = 10 ps with the excitation energy of 2.0 eV (above-gap) and density of 4.5 x 10¹² cm⁻². The dotted lines show the $E_{THz}$ waveform. (f) THz photoconductivity dynamics as a function of Δt for CVT (red) and flux (blue) MoSe₂. Note that the trace for flux MoSe₂ is multiplied by 10. The solid lines show the fitting curves.

The difference in the mid-gap defect densities between the two crystals is reflected in distinct photo-induced THz responses. Fig. 3d shows THz time trace, $\Delta E_{THz}$ (red curve), at $\Delta t = 10$ ps from the CVT MoSe$_2$ monolayer following above-gap excitation at $hv = 2.0$ eV at an excitation density $n_{abs} = 4.5 \times 10^{12}$ cm$^{-2}$. The sign reversal from $E_{THz}$ (grey dashed curve, without pump) reflects the positive photoconductivity also observed in the other CVT monolayers of WS$_2$ (Fig. 1c) and WSe$_2$ (shown below in Fig. 4). In stark contrast, the $\Delta E_{THz}$ from the flux MoSe$_2$ monolayer, blue curve in Fig. 3e, is small but of the same sign as $E_{THz}$ (grey dashed curve), indicating that the transmitted THz field is increased by photoexcitation, i.e., negative photoconductivity. The negative photoconductivity can be attributed to the formation of exciton complexes[26], i.e., exciton polarons commonly referred to as trions[42–44]. The MoSe$_2$ monolayer with low defect density from flux grown crystals is known to be n-doped; upon photo-excitation, those free electrons can combine with photogenerated excitons to create negative trions[45]. Without photo-excitation, the intrinsically doped electrons in the conduction band absorbs THz radiation and reduces the transmitted THz field. Upon photo-excitation, trion formation leads to an increased carrier effective mass, thus reducing THz absorption[26]. The negative photoconductivity, represented by $-\Delta E_{THz}/E_{THz}$ (blue dots in Fig. 3f) in the flux MoSe$_2$ monolayer is characterized by a single exponential recovery (blue fit) with a time constant ~20 ps, in agreement with the trion recombination time[26]. The positive photoconductivity in the CVT monolayer (red dots in Fig. 3f) decays on a similar time scale (~ 20 ps), but the exponential fit (red curve) reveals a residual signal, i.e., longer lived carriers attributed to exciton dissociation. These results show that the defect-mediated carrier generation process existing in the CVT monolayers is suppressed in flux grown monolayers. In the flux grown monolayer with much lower mid-gap trap density in the $10^{-9}$ cm$^{-2}$ region, above gap excitation results in intrinsic exciton physics. See also Supplementary Fig. S7 comparing the amplitudes of photoconductivity of CVT and flux MoSe$_2$.

The correlation of spontaneous exciton dissociation with mid-gap defect states is also confirmed in WSe$_2$ monolayers from CVT and flux grown WSe$_2$ crystals, with defect density from the former being one order of magnitude higher than that in the latter [14,41]. Fig. 4a and 4b show THz time traces, $\Delta E_{THz}$, at $\Delta t$ = 20 ps from CVT (red) and flux (green) WSe$_2$ monolayers, respectively, following above gap excitation at h$\nu$ = 2.30 eV and at excitation density $n_{abs}$ = 5.0 x 10$^{12}$ cm$^{-2}$. The THz photoconductivity at $\Delta t$ = 20 ps is 8x lower in the flux WSe$_2$ monolayer (Fig. 4b) than that in the CVT monolayer, Fig. 4a. Interestingly, this difference between the two samples varies strongly with $\Delta t$. Fig. 4c shows the THz photoconductivity, -$\Delta E_{THz}/E_{THz}$, as a function of delay time for both samples. The -$\Delta E_{THz}/E_{THz}$ values from CVT and flux WSe$_2$ monolayers are nearly the same immediately following photo-excitation but diverge to a ratio of ~8 at $\Delta t \geq$ 20 ps. Specifically, -$\Delta E_{THz}/E_{THz}$ from flux WSe$_2$ monolayer decays faster on time scales of 0.5±0.1 ps and 5±1 ps (bi-exponential fit as solid grey curve). For comparison, the signal from CVT monolayer decays on much longer time scales of 1.9±0.2 ps and 20±2 ps (bi-exponential fit as dashed grey curve). Because the THz probe predominantly detects charge carriers, this result suggests that the population of charge carriers decays more efficiently in the flux WSe$_2$ monolayer than the more defective CVT monolayer.

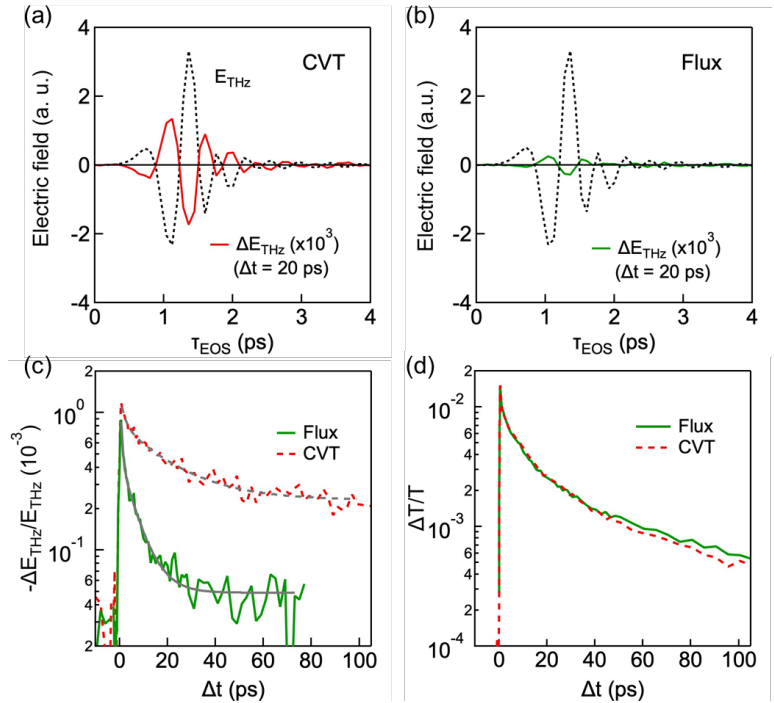

**Fig. 4 | Intrinsic exciton formation in flux WSe$_2$.** (a,b) Photoinduced THz transients (solid lines) for CVT and flux WSe$_2$ monolayers, respectively, at $\Delta t$ = 20 ps with the excitation energy of 2.30 eV and density of 5.0 x 10$^{12}$ cm$^{-2}$. The dotted lines show the $E_{THz}$ waveform. (c) THz conductivity dynamics for CVT (dotted) and flux (solid) WSe$_2$ monolayers, showing a faster conductivity decay in flux WSe$_2$ due to the formation of charge-neutral excitons. The grey curves are fitting results. (d) TA bleaching dynamics monitored at the 1s A-exciton resonance of the same CVT and flux monolayer WSe$_2$ as THz photoconductivity was measured under the same excitation condition.

The results in Fig. 4c appear counter-intuitive. To understand the origin, we used transient absorption (TA) spectroscopy since it can measure the total population of charge carriers *and* charge-neutral excitons from the bleaching of the 1s A-exciton resonance (see Supplementary Fig. S8) [30,46]. Fig. 4d compares photoinduced bleach ($\Delta T/T$, photo-induced transmission normalized by transmission without excitation), integrated over the 1s exciton resonance in CVT (red) and flux (green) samples. We use the same excitation condition, $h\nu = 2.30$ eV and $n_{abs} = 5.0$ x $10^{12}$ cm$^{-2}$, for both monolayers. The $\Delta T/T$ time profiles are essentially the same and they deviate from each other only slightly at longer times. Thus, the presence of mid-gap defect states does not contribute appreciably to the overall decay dynamics of electronic excitations (see Supplementary Note 4 for more discussion of bleaching amplitudes due to excitons and carriers[32,47]). Since the TA measurement establishes similar dynamics in both CVT and flux WSe$_2$ monolayers (Fig. 4d), the much faster decay of THz photoconductivity in the latter (Fig. 4c) cannot be due to the decay of the overall electronic excitation. Rather, we attribute the fast decay of THz photoconductivity in the flux WSe$_2$ monolayer on the 0.5-5 ps time scale to a change in the identity of the photo-excited species, particularly the formation of insulating excitons from the unbound and conducting electrons and holes formed initially from above gap excitation[38,48]. With this in mind, the long-lived THz signal in CVT monolayers is attributed to a substantial portion of the electronic excitation remaining as carriers, in contrast to the common view of the dominance of excitons.

<u>Defect-mediated exciton dissociation: an Auger interpretation</u>

The THz photoconductivity in CVT monolayer TMDCs is independent of temperature, is correlated with mid-gap trap states, and is observed at sufficiently low excitation density. In a defect-mediated process involving mid-gap state at the single excitation limit, the requirement for overcoming the exciton binding energy can be realized in an Auger scattering mechanism, as discussed theoretically before for nonradiative recombination[39,40]. Fig. 5 illustrates the proposed intra-exciton Auger mechanism for exciton dissociation in a TMDC monolayer. We adopt the rigorous energy level diagram based on either ionization energy or electron affinity for both carriers and excitons, Fig. 5.[49] The energy liberated from hole trapping by a mid-gap state can satisfy the energy requirement for exciton dissociation which releases the free electron into the conduction band in an Auger scattering process, Fig. 5a. The same argument applies to electron trapping and releasing the free hole to the valence band, Fig. 5b. Recent theoretical work suggests that Auger scattering processes are particularly efficient for monolayer TMDCs because the same strong Coulomb potential responsible for large exciton binding energies also leads to high electron/hole trapping and high Auger scattering rates[39,40]. The estimated electron/hole trapping time for an exciton in a TMDC monolayer is $\leq 1\mathrm{ps}$.[39] This is the same time scale for the Auger mechanism to release a free electron or hole, in agreement with the observed ultrafast rise in THz photoconductivity.

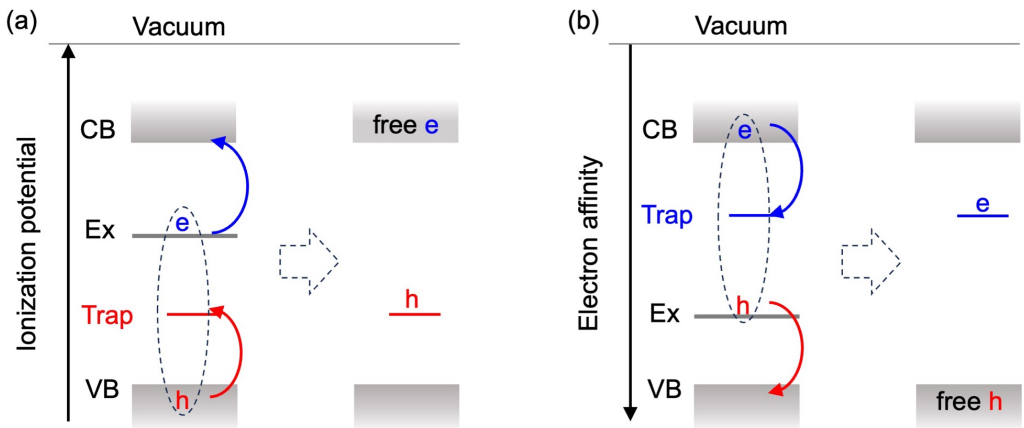

**Fig. 5. | Defect-mediated exciton dissociation.** (a) Drawn on the energy scale of ionization potential, the energy of hole trapping to a mid-gap state can satisfy the energy requirement for dissociating the exciton and releasing a free electron to the conduction band. (b) Similar diagram for electron trapping and hole releasing on an electron affinity scale. CB (VB): conduction (valence) band. Ex: exciton level.

<u>Summary</u>

Our findings have solved a major conundrum in the photophysics of TMDC monolayers, namely the seemingly contradicting results on large exciton binding energies and free carrier generation in various optoelectronic devices. These findings also suggest that the common assumption of excitons being the dominant quasiparticle formed in photoexcited TMDC monolayers is only valid in the limit of low defect densities. The fact that ~10% of excitons spontaneously dissociate into carriers in photo-excited TMDC monolayers exfoliated from commonly available CVT crystals indicates that one needs to be cautious when applying the widely held views of exciton physics, especially when defect densities are unknown. The poor screening at the 2D limit is not only responsible for the large exciton binding energy but also enhances any process involving the Coulomb potential, as suggested here for the trap-mediated Auger scattering and exciton dissociation. While reducing defect density to reach the intrinsic physical limit is always an important goal in 2D materials research, one may also intentionally incorporate or control these defects for efficient carrier generation in optoelectronics, such as photodetectors.

**Methods**

**Crystal growth of self-flux crystals**

We synthesized MoSe$_2$ and WSe$_2$ crystals using a self-flux method with excess Se [14,50]. Mo (W) powder of 99.997% (99.999%) purity was loaded into a quartz ampule with Se shot of 99.999+% purity in a metal-to-chalcogen molar ratio ranging from 1:5 to 1:100. The ampule was evacuated to ~$10^{-5}$ torr and sealed. The sealed ampule was heated to 1000˚C over 24 hours, held at that temperature for 2 weeks, then cooled at a rate of 1˚C/hr to 500˚C, at which the cooling rate was increased to 5˚C/hr down to room temperature (rt). The ampule was then reheated back to 1000˚C over 24 hrs, held at that temperature for 5 days, and the same cooling procedure was done down to room temperature. The TMD and excess selenium contents were transferred to a new quartz ampule, and a piece of quartz wool was added and pushed down to about 1 cm above the TMD crystals and excess selenium. The second ampule was sealed under vacuum at the aforementioned pressure, then heated to 285˚C to melt the unreacted Se and dwelled for 30 min, then removed from the furnace, flipped and centrifuged. The TMD crystals were collected from the quartz wool and transferred to a third ampule, then sealed under vacuum. The crystals were annealed for 24 hrs in a temperature gradient with the crystals at the hot end at 285˚C, and the cold end of the ampule held at room temperature. The purpose of this this annealing step was to melt and remove any excess selenium from the crystals.

**Gold-tape exfoliation of large-area monolayer TMDs**

Macroscopic TMD monolayers were exfoliated via the gold tape method[13]. Bulk WS$_2$, WSe$_2$, and MoSe$_2$, grown via chemical vapor transport (CVT), were purchased from HQ Graphene; flux WSe$_2$ and MoSe$_2$ crystals were grown as described above. We prepared the tapes by evaporating gold onto a polished silicon wafer before spin coating a protective layer of polyvinylpyrrolidone (PVP). We deposited gold onto the polished silicon wafer at a rate of 0.05 nm/s (Angstrom Engineering EvoVac Multi-Process thin film deposition system). The PVP protection layer was prepared from spinning a solution of PVP (40,000mw Alpha Aesar), ethanol, and acetonitrile with a 2:9:9 mass ratio onto the gold surface of the wafer (1000 rpm, 1000 rpm/s acceleration, 2 min) before curing on a hot plate (150 °C, 5 min). Once prepared, the gold tape was removed from the silicon wafer by thermal release tape (Semiconductor Equipment Corp. Revalpha RA-95LS(N)). The gold tape was gently pressed onto the surface of a bulk TMD crystal to exfoliate a large area monolayer, which was subsequently transferred onto a 0.5 mm thick z-cut quartz substrate (MTI Corporation). The sample assembly was heated on a hot plate at 130 °C until the thermal release tape went cloudy and could be easily removed. The sample assembly was soaked in deionized water for 3 hrs to remove the PVP protection layer and in acetone for 1 hr to remove any remaining polymer residue. Then, the gold was dissolved with a 5-minute bath in a KI/$I_2$ gold etchant solution (Iodine, 99.99%, Alfa Aesar; potassium iodide, 99.0%, Sigma-Aldrich, and deionized water with a 4:1:40 mass ratio). The sample was soaked in deionized water for 2 hours to remove any residual etchant, rinsed in isopropanol, and dried with $N_2$.

**Optical pump-THz probe spectroscopy**

The setup for optical-pump/THz-probe spectroscopy is schematically illustrated in Supplementary Fig. S9. For the light source, we used a Ti:sapphire regenerative amplifier (RA) with a pulse duration of 30 fs, repetition rate of 10 kHz, and wavelength centered at 800 nm (Coherent, Legend). The RA output was split into three beams for THz generation, THz sampling, and optical pump, respectively. The THz-probe pulse was generated in a two-color air plasma method, where the fundamental 800 nm beam passes through a beta barium borate (BBO) crystal to generate the frequency doubled beam. Both the fundamental and frequency-doubled beams were are focused onto the same point in air generating a broadband THz probe pulse. The THz beam was collimated by an off-axis parabolic mirror and transmitted through a high-resistivity Si wafer, which blocks the fundamental and frequency-doubled beams while transmitting the THz field. The THz beam was then focused onto a sample mounted in a $LN_2$-flow-cryostat, after which it was collimated by another parabolic mirror. High-density polyethylene (HDPE) and Teflon polymers plates were used to filter the optical pump beam while allowing the transmission of the THz probe. The THz probe was then focused onto a 1-mm (110) ZnTe crystal along with another 800-nm sampling beam, by which the real-time waveform of THz electrical field was detected via electro-optic sampling (EOS). Following the ZnTe crystal, the sampling beam passes through a quarter-wave plate (QWP) and a Wollaston prism, and the split sampling beams were detected on a balance detector (Thorlabs, PDB210A). The amplified signal from the balance detector was integrated with a boxcar integrator (Stanford Research Systems, SR250) and the output was recorded using a data acquisition (DAQ) card (National Instruments, USB-6216).

The femtosecond optical pump beam was generated using a home-built non-collinear parametric amplifier (NOPA) pumped by the second harmonic of the Ti:sapphire laser. The center wavelengths were tuned in the range of 500-620 nm with enough fluence for the large-area far-field THz measurements. The beam spot sizes at the sample position were determined with a knife-edge method using a photodetector (Thorlabs, DET100A2) and pyroelectric terahertz power detectors (Gentec-EO, THZ9B-BL-DA) equipped with lock-in amplifier (Stanford Research Systems, SR830) for optical pump and THz probe, respectively. In front of the THz power detector, additional polymer filters were used to block the higher-frequency THz beam from entering the detector. The typical spot sizes (1/e diameter in power) were 2 mm and 0.15 mm for optical pump and THz probe, respectively. The absorbed photon density was determined via $n_{abs} = a_{eff}P_{ex} \times (fAh\nu)^{-1}$, where $a_{eff}$ is the effective absorptivity, $P_{ex}$ the laser power, $f = 5$ kHz the effective repetition rate after the chopper, $A$ the beam area, and $h\nu$ the photon energy. The effective absorptivity was calculated from the convolution of absorptivity spectrum $a(E)$ and pump spectrum $I_{pump}(E)$ $a_{eff} = \left[\int dE\, a(E) \times I_{pump}(E)\right]/\left[\int dE\, I_{pump}(E)\right]$. $a_{eff}$ was 0.049 (0.031) for the above gap (resonant) excitation of $WS_2$, 0.072 for $WSe_2$, and 0.043 for $MoSe_2$. We optimized the spatial overlap between the optical pump and THz probe by inserting a Si wafer as a reference (which gave strong signal), and then replaced it with the actual sample. The Si wafer was also used to confirm the sign of photoinduced signal $\Delta E_{THz}$. The relatively high repetition rate of 10 kHz together with the DAQ acquisition scheme[33], the use of filters to block optical beams, and the optimization of electronics allowed for high signal to noise ration. We confirmed that there was no photoinduced THz signal from the $SiO_2$ substrate by measuring the area without the TMD monolayer (Supplementary Fig. 10).

**Scanning tunneling microscopy/spectroscopy**

STM measurements were performed on a commercial Omicron ultra-high vacuum STM system. Single crystals of $MoSe_2$ and $WSe_2$ were mounted onto metallic sample holders by vacuum compatible silver paste (EPO-TEK H20E). Samples were then exfoliated to expose a clean surface and transferred into the STM chamber. A chemically etched tungsten STM tip was conditioned and calibrated on an Au(111) single crystal prior to the measurements. Measurements were performed at room temperature (300 K).

**Photoinduced conductivity spectra**

The time-domain THz electric fields obtained from EOS were converted to the frequency-domain spectra $E_{THz}(\omega)$ by Fourier transformation. Complex spectra of the photoinduced conductivity were then obtained under thin film approximation: $\Delta\sigma(\omega) = -\frac{1+n_{sub}}{Z_0}\frac{\Delta E_{THz}(\omega)}{E_{THz}(\omega)}$. Here, $\Delta E_{THz}(\omega) = E_{THz,w/\,pump}(\omega) - E_{THz,w/o\,pump}(\omega)$ is the photoinduced change in the THz field, $Z_0 = 377\,\Omega$ is the vacuum impedance, and $n_{sub}$ is the complex refractive index of the $SiO_2$ substrate. $n_{sub}$ was determined independently by performing the THz spectroscopy on the substrate: $n_{sub} = 2.19 + 0i$.

We analyzed the complex photoconductivity spectra $\Delta\sigma(\omega)$ with the Drude-Smith model[36], $\Delta\sigma = \frac{ne^2\tau/m_{eff}}{1-i\omega\tau}\left[1 + \frac{c_{DS}}{1-i\omega\tau}\right]$, where $n$ is the carrier density, $e$ the elementary charge, $m_{eff}$ the effective mass, $\tau$ the scattering time, $\omega$ the angular frequency, $c_{DS}$ the Smith parameter. For the fitting, the effective mass $m_{eff} = (m_e^{-1} + m_h^{-1})^{-1}$ is the average of the electron and hole effective masses reported in Ref. [37]. The real and imaginary parts were fit simultaneously with a global fitting procedure.

## References

1. Mak, K. F., Lee, C., Hone, J., Shan, J. & Heinz, T. F. Atomically Thin MoS2: A New Direct-Gap Semiconductor. *Phys Rev Lett* **105**, 136805 (2010).

2. Splendiani, A. *et al.* Emerging Photoluminescence in Monolayer MoS2. *Nano Lett* **10**, 1271–1275 (2010).

3. Wang, G. *et al.* Colloquium : Excitons in atomically thin transition metal dichalcogenides. *Rev Mod Phys* **90**, 021001 (2018).

4. Wilson, N. P., Yao, W., Shan, J. & Xu, X. Excitons and emergent quantum phenomena in stacked 2D semiconductors. *Nature* **599**, 383–392 (2021).

5. Lee, C.-H. *et al.* Atomically thin p–n junctions with van der Waals heterointerfaces. *Nat Nanotechnol* **9**, 676–681 (2014).

6. Klots, A. R. *et al.* Probing excitonic states in suspended two-dimensional semiconductors by photocurrent spectroscopy. *Sci Rep* **4**, 6608 (2014).

7. Furchi, M. M., Pospischil, A., Libisch, F., Burgdörfer, J. & Mueller, T. Photovoltaic Effect in an Electrically Tunable van der Waals Heterojunction. *Nano Lett* **14**, 4785–4791 (2014).

8. Baugher, B. W. H., Churchill, H. O. H., Yang, Y. & Jarillo-Herrero, P. Optoelectronic devices based on electrically tunable p–n diodes in a monolayer dichalcogenide. *Nat Nanotechnol* **9**, 262–267 (2014).

9.   Chen, P. *et al.* Approaching the intrinsic exciton physics limit in two-dimensional semiconductor diodes. *Nature* **599**, 404–410 (2021).

10.  Haastrup, S., Latini, S., Bolotin, K. & Thygesen, K. S. Stark shift and electric-field-induced dissociation of excitons in monolayer MoS2 and hBN/MoS2 heterostructures. *Phys Rev B* **94**, 041401 (2016).

11.  Kaindl, R. A., Hägele, D., Carnahan, M. A. & Chemla, D. S. Transient terahertz spectroscopy of excitons and unbound carriers in quasi-two-dimensional electron-hole gases. *Phys Rev B* **79**, 045320 (2009).

12.  Ulbricht, R., Hendry, E., Shan, J., Heinz, T. F. & Bonn, M. Carrier dynamics in semiconductors studied with time-resolved terahertz spectroscopy. *Rev Mod Phys* **83**, 543–586 (2011).

13.  Liu, F. *et al.* Disassembling 2D van der Waals crystals into macroscopic monolayers and reassembling into artificial lattices. *Science (1979)* **367**, 903–906 (2020).

14.  Edelberg, D. *et al.* Approaching the Intrinsic Limit in Transition Metal Diselenides via Point Defect Control. *Nano Lett* **19**, 4371–4379 (2019).

15.  Zhang, L. *et al.* Van der Waals heterostructure polaritons with moiré-induced nonlinearity. *Nature* **591**, 61–65 (2021).

16.  Gu, J. *et al.* Enhanced nonlinear interaction of polaritons via excitonic Rydberg states in monolayer WSe2. *Nat Commun* **12**, 2269 (2021).

17.  Regan, E. C. *et al.* Emerging exciton physics in transition metal dichalcogenide heterobilayers. *Nat Rev Mater* **7**, 778–795 (2022).

18.  Mak, K. F. & Shan, J. Photonics and optoelectronics of 2D semiconductor transition metal dichalcogenides. *Nat Photonics* **10**, 216–226 (2016).

19.  Yang, R. *et al.* 2D Transition Metal Dichalcogenides for Photocatalysis. *Angewandte Chemie* **135**, e202218016 (2023).

20.  Wang, Z. *et al.* Evidence of high-temperature exciton condensation in two-dimensional atomic double layers. *Nature* **574**, 76–80 (2019).

21.  Zeng, Y. *et al.* Exciton density waves in Coulomb-coupled dual moiré lattices. *Nat Mater* **22**, 175–179 (2023).

22.  Xu, Y. *et al.* Correlated insulating states at fractional fillings of moiré superlattices. *Nature* **587**, 214–218 (2020).

23.  Bae, Y. J. *et al.* Exciton-Coupled Coherent Magnons in a 2D Semiconductor. *Nature* **608**, 282–286 (2022).

24.  Shimazaki, Y. *et al.* Strongly correlated electrons and hybrid excitons in a moiré heterostructure. *Nature* **580**, 472–477 (2020).

25. Docherty, C. J. *et al.* Ultrafast Transient Terahertz Conductivity of Monolayer MoS 2 and WSe 2 Grown by Chemical Vapor Deposition. *ACS Nano* **8**, 11147–11153 (2014).

26. Lui, C. H. *et al.* Trion-Induced Negative Photoconductivity in Monolayer MoS2. *Phys Rev Lett* **113**, 166801 (2014).

27. Gustafson, J. K., Cunningham, P. D., McCreary, K. M., Jonker, B. T. & Hayden, L. M. Ultrafast Carrier Dynamics of Monolayer WS 2 via Broad-Band Time-Resolved Terahertz Spectroscopy. *The Journal of Physical Chemistry C* **123**, 30676–30683 (2019).

28. Xu, S., Yang, J., Jiang, H., Su, F. & Zeng, Z. Transient photoconductivity and free carrier dynamics in a monolayer WS 2 probed by time resolved Terahertz spectroscopy. *Nanotechnology* **30**, 265706 (2019).

29. Siday, T. *et al.* Ultrafast Nanoscopy of High-Density Exciton Phases in WSe 2. *Nano Lett* **22**, 2561–2568 (2022).

30. Wang, J. *et al.* Optical generation of high carrier densities in 2D semiconductor heterobilayers. *Sci Adv* **5**, 2–10 (2019).

31. Steinhoff, A. *et al.* Exciton fission in monolayer transition metal dichalcogenide semiconductors. *Nat Commun* **8**, 1166 (2017).

32. Chernikov, A., Ruppert, C., Hill, H. M., Rigosi, A. F. & Heinz, T. F. Population inversion and giant bandgap renormalization in atomically thin WS2 layers. *Nat Photonics* **9**, 466–469 (2015).

33. Werley, C. A., Teo, S. M. & Nelson, K. A. Pulsed laser noise analysis and pump-probe signal detection with a data acquisition card. *Review of Scientific Instruments* **82**, 123108 (2011).

34. Li, Y. *et al.* Measurement of the optical dielectric function of monolayer transition-metal dichalcogenides: MoS2, MoSe2, WS2, and WSe2. *Phys Rev B* **90**, 205422 (2014).

35. Chernikov, A. *et al.* Exciton Binding Energy and Nonhydrogenic Rydberg Series in Monolayer WS2. *Phys Rev Lett* **113**, 076802 (2014).

36. Smith, N. Classical generalization of the Drude formula for the optical conductivity. *Phys Rev B* **64**, 155106 (2001).

37. Kormányos, A. *et al.* k.p theory for two-dimensional transition metal dichalcogenide semiconductors. *2d Mater* **2**, 022001 (2014).

38. Steinleitner, P. *et al.* Direct Observation of Ultrafast Exciton Formation in a Monolayer of WSe 2. *Nano Lett* **17**, 1455–1460 (2017).

39. Wang, H. *et al.* Fast exciton annihilation by capture of electrons or holes by defects via Auger scattering in monolayer metal dichalcogenides. *Phys Rev B* **91**, 165411 (2015).

40. Wang, H., Zhang, C. & Rana, F. Ultrafast Dynamics of Defect-Assisted Electron–Hole Recombination in Monolayer MoS2. *Nano Lett* **15**, 339–345 (2015).

41.    Rhodes, D., Chae, S. H., Ribeiro-Palau, R. & Hone, J. Disorder in van der Waals heterostructures of 2D materials. *Nat Mater* **18**, 541–549 (2019).

42.    Efimkin, D. K. & MacDonald, A. H. Many-body theory of trion absorption features in two-dimensional semiconductors. *Phys Rev B* **95**, 035417 (2017).

43.    Sidler, M. *et al.* Fermi polaron-polaritons in charge-tunable atomically thin semiconductors. *Nat Phys* **13**, 255–261 (2017).

44.    Rana, F., Koksal, O. & Manolatou, C. Many-body theory of the optical conductivity of excitons and trions in two-dimensional materials. *Phys Rev B* **102**, 85304 (2020).

45.    Kim, B. *et al.* Free Trions with Near-Unity Quantum Yield in Monolayer MoSe2. *ACS Nano* **16**, 140–147 (2022).

46.    Ceballos, F., Cui, Q., Bellus, M. Z. & Zhao, H. Exciton formation in monolayer transition metal dichalcogenides. *Nanoscale* **8**, 11681–11688 (2016).

47.    Schmitt-Rink, S., Chemla, D. S. & Miller, D. a B. Theory of transient excitonic optical nonlinearities in semiconductor quantum-well structures. *Phys Rev B* **32**, 6601–6609 (1985).

48.    Ceballos, F., Cui, Q. N., Bellus, M. Z. & Zhao, H. Exciton formation in monolayer transition metal dichalcogenides. *Nanoscale* **8**, 11681–11688 (2016).

49.    Zhu, X. -Y. How to draw energy level diagrams in excitonic solar cells. *Journal of Physical Chemistry Letters* **5**, 2283–2288 (2014).

50.    Liu, S. *et al.* Two-step flux synthesis of ultrapure transition metal dichalcogenides. *arXiv preprint arXiv:2303.16290* (2023).

**Acknowledgements.**

Sample preparation and characterization were supported by the Materials Science and Engineering Research Center (MRSEC) through NSF grant DMR-2011738. The THz spectroscopic measurement is supported in part by the US Army Research Office, grant number W911NF-23-1-0056. XYZ and MB acknowledge support for collaboration by the Max Planck – New York City Center for Non-Equilibrium Quantum Phenomena. T.H. acknowledges support by JSPS Overseas Postdoctoral Research Fellowship program.

**Data Availability.**

The data represented in Figs. 1-4 are provided with the article source data. All data that support the results in this article are available from the corresponding author upon reasonable request.

**Author contributions.**

TH and XYZ conceived this work. TH carried out optical measurements with assistance by LH. MAH performed the STM and STS under the supervision by ANP and JCH. LNH synthesized the flux grown crystals under the supervision by KB and JCH. NO exfoliated the large area monolayers. TH, XYZ, HIW, and MB analyzed the data. XYZ supervised the project. The manuscript was prepared by TH and XYZ in consultation with all other authors. All authors read and commented on the manuscript.

**Competing Interests.**

All authors declare that they have no competing interests

**Supplementary Information for**

**Spontaneous Exciton Dissociation in Transition Metal Dichalcogenide Monolayers**

Taketo Handa[1], Madisen A. Holbrook[2], Nicholas Olsen[1], Luke N. Holtzman[3], Lucas Huber[1], Hai I. Wang[4], Mischa Bonn[4], Katayun Barmak[3], James C. Hone[5], Abhay N. Pasupathy[2], X.-Y. Zhu[1,†]

**Contents:**

**References**

## Note 1. Contribution of excitons to low-frequency THz conductivity spectra

We estimate the contributions of excitons on the THz signals for the present 1-3 THz region, by which we can confirm its negligible contribution due to the large exciton binding energy. We estimate the conductivity of excitons based on the reported dielectric function assessed with the mid infrared (MIR) spectroscopy[1]. Using MIR light, the Rydberg transition between intraexcitonic 1s-2p states was directly measured for monolayer WSe$_2$. The result was well explained with a simple Lorentz oscillator model:

$$\Delta\epsilon_{Ex}(\omega) = \frac{n_{Ex}e^2}{d\epsilon_0\mu} \frac{f_{1s-2p}}{\frac{E_{res}}{\hbar^2} - \omega^2 - i\omega\Delta},$$

where $\omega$ represents the angular frequency, $n_{Ex}$ the photogenerated exciton density, $d$ effective thickness of the monolayer, $\mu$ the exciton reduced mass, $f_{1s-2p}$ oscillator strength of the transition, $E_{res}$ the transition energy, $\Delta$ the damping term. $e$, $\epsilon_0$, and $\hbar$ are the elementary charge, vacuum permittivity, and reduced Planck constant, respectively. In Ref. [1], $f_{1s-2p}$ was estimated theoretically to be 0.32, and $\mu = 0.17m_0$ is the reduced mass for WSe$_2$ with $m_0$ being the bare electron mass. Through the MIR experiment, $E_{res}$ was determined to be 167 meV with the peak width of $\Delta = 99$ meV.

For our monolayer WS$_2$, we use essentially the same parameters with the Lorentz oscillator equation, while we use the following values for WS$_2$, $\mu = 0.149m_0$[2] and $E_{res} = 163$ meV estimated from the reflectance measurement on monolayer WS$_2$ on SiO$_2$[3]. With $n_{Ex} = 10^{12}$ cm$^{-2}$ and the conversion to the conductivity via $\Delta\sigma_1 = \omega\epsilon_0\epsilon_2$ and $\Delta\sigma_2 = -\omega(\epsilon_0\epsilon_1 - \epsilon_0)$, we obtain Supplementary Fig. 2. The result reveals that the contribution of excitons on the conductivity at the present THz region of 2THz is $\Delta\sigma_1 = 3.8 \times 10^{-9}$ ($\Omega^{-1}$) and $\Delta\sigma_2 = -4.6 \times 10^{-8}$ ($\Omega^{-1}$). These are much smaller than our detected signals, showing the negligible contribution.

A separate check can be done using the reported polarizability of excitons. The polarizability for monolayer WS$_2$ surrounded by the effective dielectric constant of 1.5 (air and SiO$_2$) is $6 \times 10^{-18}$ (eV m$^{-2}$ V$^{-2}$)[4]. By using $n_{Ex} = 10^{12}$ cm$^{-2}$ and neglecting geometrical correction, $\Delta\sigma_2 = -1.2 \times 10^{-7}$ ($\Omega^{-1}$), again much smaller than the experimental results, e.g., see Fig. 1e in the main text.

## Note 2. Thermodynamic ionization ratio

Here, we calculate the thermodynamic ionization ratio of excitons into unbound electrons and holes considering thermodynamic equilibrium between them. Upon photoexcitation, optically generated e-h pairs may form tightly-bound excitons. These excitons may also ionize into electrons and holes. This ionization process can be well understood using the ionization ratio $\alpha = n_c/(n_c + n_x)$, where $n_x$ and $n_c$ represent the exciton and charge carrier density, respectively. First, we focus on moderate to weak excitation densities ($10^9$-$10^{12}$ cm$^{-2}$) below the exciton Mott transition density, which is relevant for most of optical studies and applications. In this regime, the thermodynamic mass action law governs the system with the Saha equation[5]: $\frac{\alpha}{1-\alpha} = \frac{k_B T}{2\pi\hbar^2(n_c+n_X)}\mu e^{-E_b/k_B T}$, where $E_b$ represents the exciton binding energy, $k_B T$ the thermal energy, $\mu$ the reduced mass. By using the reported $E_b$ = 320 meV[3] for monolayer WS$_2$ on SiO$_2$ and $\mu$ = 0.17[2], we calculate the ionization ratio for different temperatures in Supplementary Fig. 5. At the relevant excitation density of $10^{12}$ cm$^{-2}$, $\alpha$ is only 0.001 for 293 K, confirming little charge carrier population and the dominance of excitons under the thermodynamic consideration. At lower temperature, the thermal dissociation is significantly suppressed, predicting almost complete excitonic picture.

When excitation density is further increased, there appears the exciton Mott transition, where the insulating excitons emerge into conductive e-h plasma phases. At around the Mott transition density $n_{Mott}$, a sharp rise in $\alpha$ is expected by theories[6] and probed by DC conductivity measurements. For monolayer TMDs, $n_{Mott}$ is around $10^{13}$-$10^{14}$ cm$^{-2}$. In Supplementary Fig. 5, we did not include the Mott transition due to its rather ambiguous nature, which will be the subject further future study.

## Note 3. Exciton diffusion and inter-exciton Auger scattering

We discuss the exciton diffusion and the effective spatial occupancy by excitons. The THz dynamics showed the rapid emergence following photoexcitation within ~1 ps, which is mostly determined by the response function of our setup. Considering the diffusion coefficient of TMDs of D ~ 1 cm$^2$/s at room temperature, excitons can diffuse ~10 nm within 1 ps. The lateral extent of excitons (twice the Bohr radius) is around 10 nm,[3] and thus, the effective spatial extent of excitons over the time can be considered as $\sqrt{(10)^2 + (1)^2} \approx 10.5$ nm. Together with the typical excitation density in the present study of $10^{12}$ cm$^{-2}$, the effective spatial occupancy within the 2D plane becomes almost unity. This ensures the overlap between the exciton wavefunction and defect wavefunction, which initiates the trap-mediated Auger-like scattering process.

Note that our excitation density dependent measurements consistently show the almost linear dependence. Based on this result, we can exclude the Auger scattering between inter-excitons as a possible origins of photoconductivity because the inter-exciton Auger process should show the dependence close to quadratic behavior.

## Note 4. Exciton bleaching amplitudes due to excitons and charge carriers

As shown in Fig. 4d, the TA bleaching dynamics monitored at the 1s A-exciton resonance are very similar for CVT and flux $WSe_2$ monolayers. The large similarity can be attributed to the fact that charge carriers in a 2D semiconductor are almost twice more efficient on inducing the bleaching than excitons due to stronger exchange and phase-space filling effects of the former.[7,8] More specifically, unbound, free electron-hole pair can induce bleaching signal as twice efficient as bound electron-hole pair (exciton) per the same electron-hole density.

We have discussed in Fig. 5 that in the presence of mid-gap states, photogenerated excitons can undergo the defect-mediated exciton dissociation process, leaving either free electron or free hole behind in a quasiparticle band. Considering the free carriers' ability of inducing the larger bleaching amplitude than excitons by factor of two, such a trapping and consequent annihilation of an exciton into only one conductive carrier would keep the total bleaching amplitude, which makes bleaching amplitude very similar for an exciton (not trapped) and dissociated exciton (i.e., free carrier).

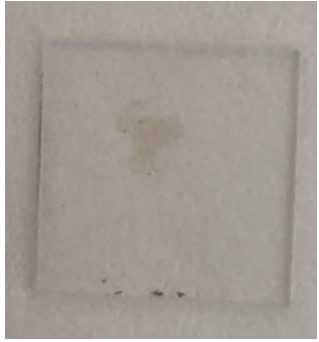

**Supplementary Fig. 1. Optical image of MoSe₂ monolayer exfoliated on a quartz substrate.**

The substrate area is 10 x 10 mm$^2$.

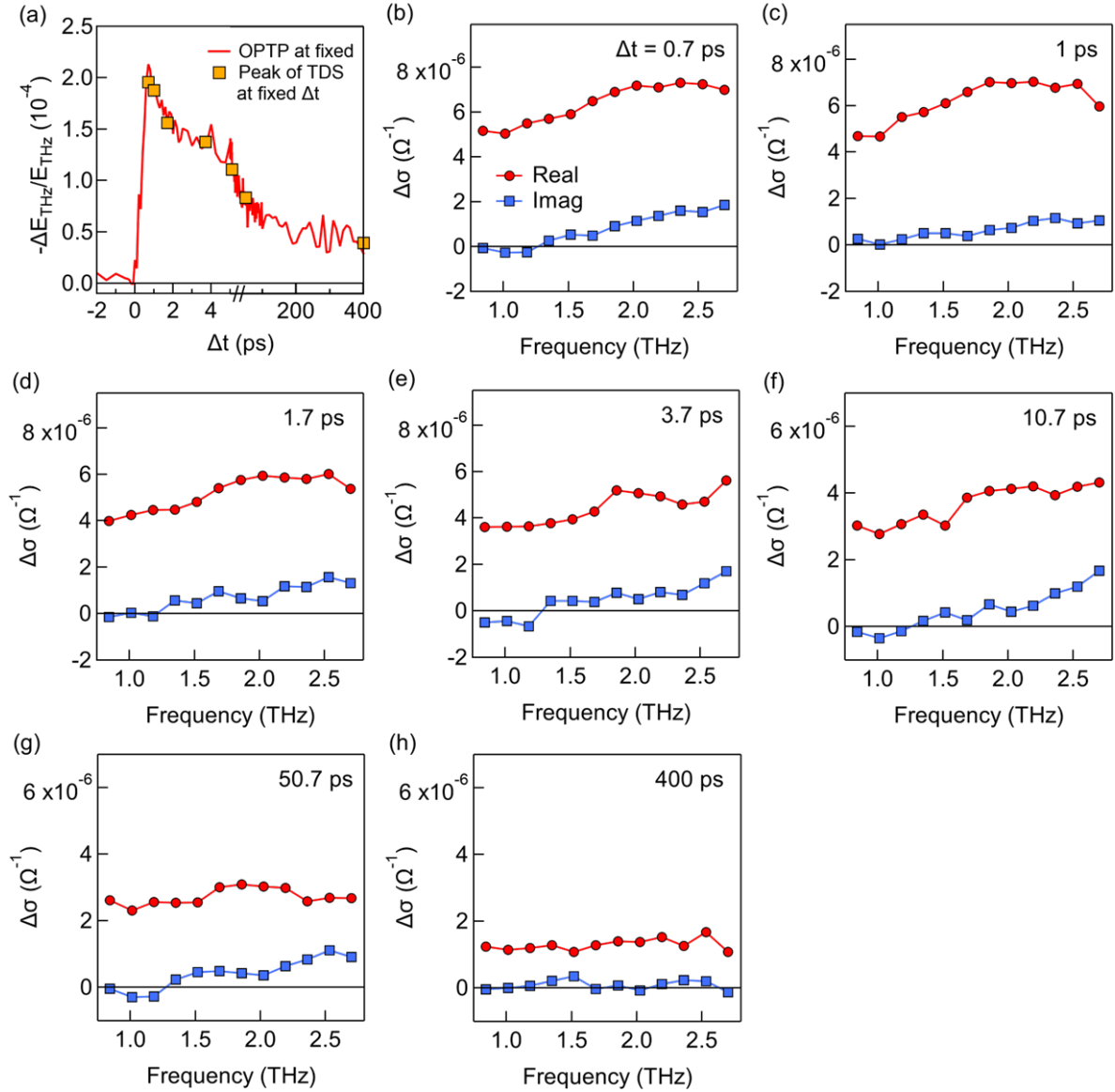

**Supplementary Fig. 2. THz conductivity dynamics and THz TDS at different pump-probe delays for the monolayer WS₂.** (a) Optical pump-THz probe (OPTP) dynamics tracked at the fixed EOS timing $\tau_{EOS}$ that gives the largest photoinduced THz signal, solid line. Also shown with the squares are the peak THz signals of THz TDS scans obtained at different pump-probe delay $\Delta t$. The condition was the above-gap excitation with the absorbed photon density of $8.5 \times 10^{11}$ cm$^{-2}$. (b-h) Complex photoconductivity spectra at different $\Delta t$, demonstrating the real part always dominates the THz response.

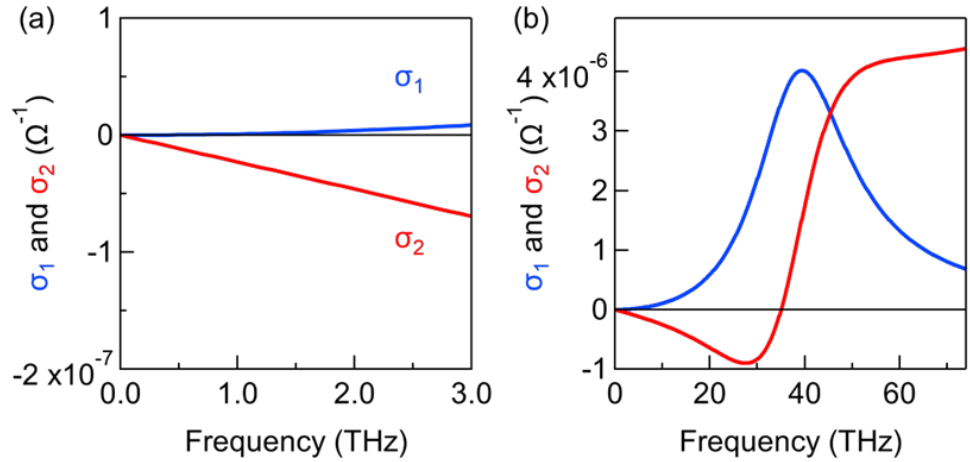

**Supplementary Fig. 3. Negligible contribution of photogenerated excitons to the THz response in the present study**. **a,b,** Calculated conductivities due to excitons with the density of $n_{Ex} = 10^{12}$ cm$^{-2}$ at the frequency range of the present study and at wider range, respectively. At the low frequency region, the THz conductivity is very small copared to the experimentally observed values due to the large exciton binding energy and consequent high resonance frequency. The detail of the calculation is provided in Supplementary Note 2.

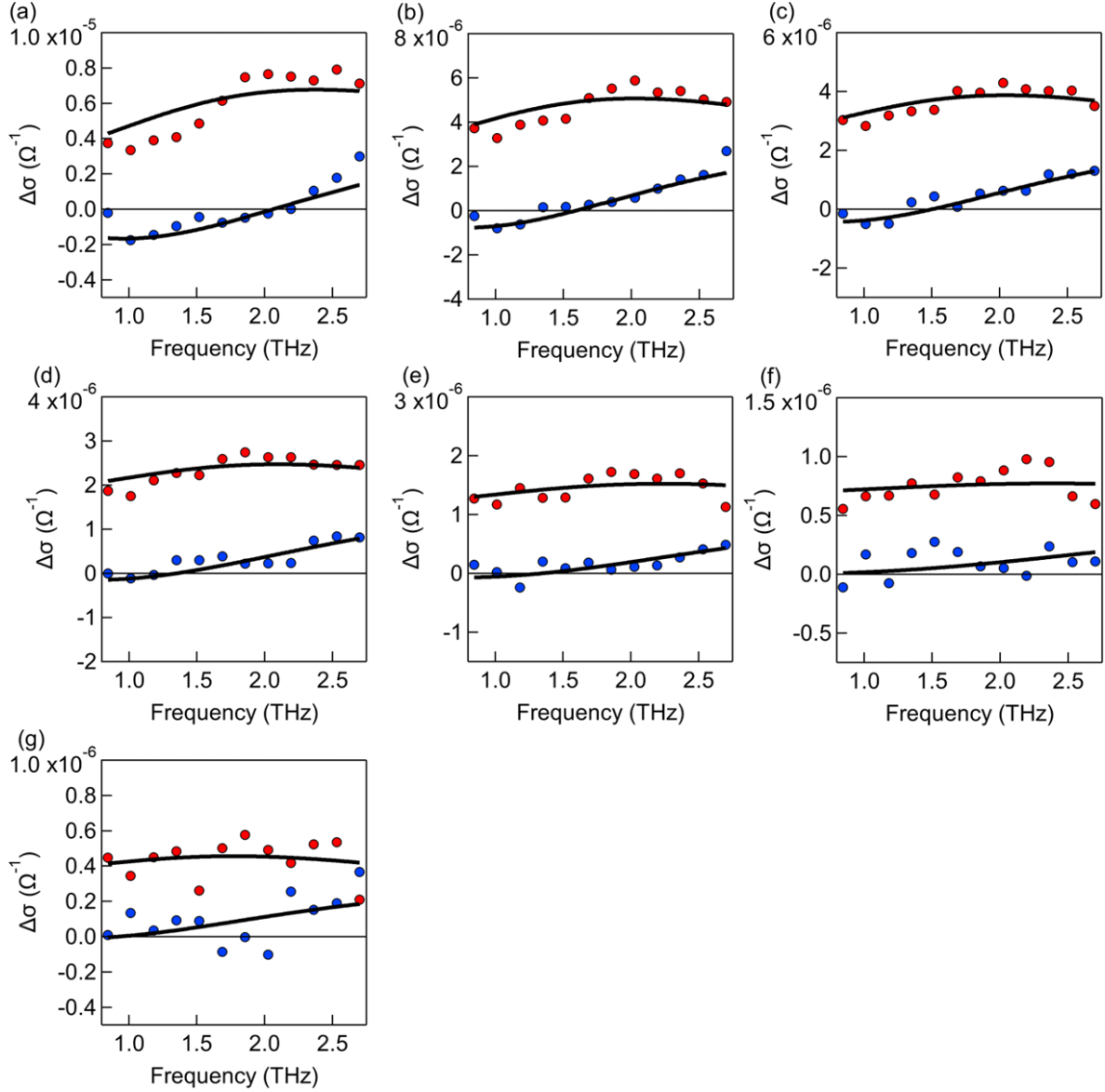

**Supplementary Fig. 4. Complex conductivity spectra for the monolayer WS₂ under the above-gap excitation at different excitation densities. a-g,** Conductivity spectra measured at Δt = 5 ps, with the absorbed photon density of $n_{abs}$ = 9.1 x 10¹², 5.2 x 10¹², 2.9 x 10¹², 1.6 x 10¹², 8.5 x 10¹¹, 4.4 x 10¹¹, 2.3 x 10¹¹, respectively. The solid lines show the fitting result to the Drude-Smith model (also see Fig. 1e in the main text).

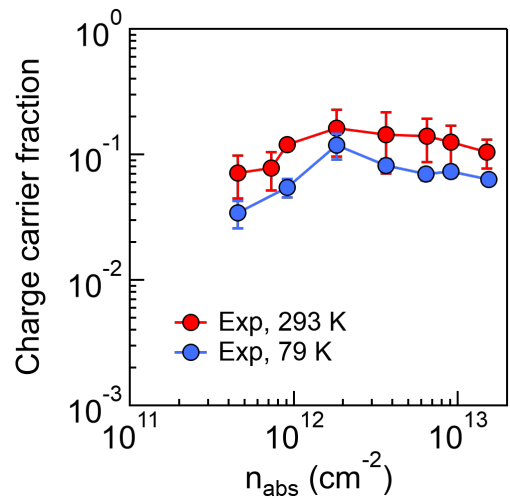

**Supplementary Fig. 5. Excitation density dependence of the charge-carrier fraction of the monolayer WS₂ on SiO₂ at 293 K and 80 K.** The corresponding excitation density dependence of the peak THz conductivity is provided in Fig. 2a in the main text. The data were taken with the above-gap excitation.

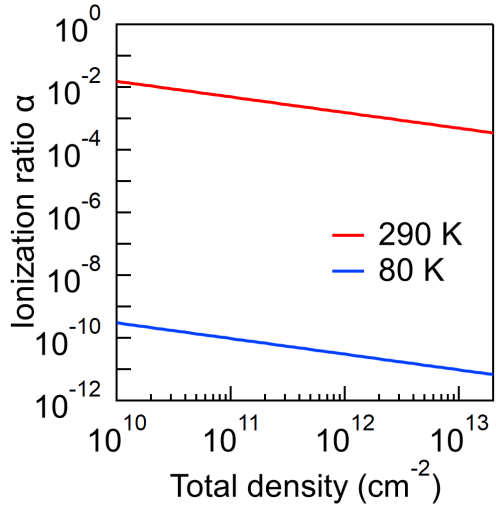

**Supplementary Fig. 6. The ionization ratio of monolayer WS₂ onSiO₂ calculated considering the thermodynamic Saha equation.** Detailed description is provided in Supplementary Note 1. The calculation shows that charge-neutral excitons are expected to be dominant when the pure thermodynamics govern the system; however, the THz TDS results revealed the rather larger charge-carrier fraction.

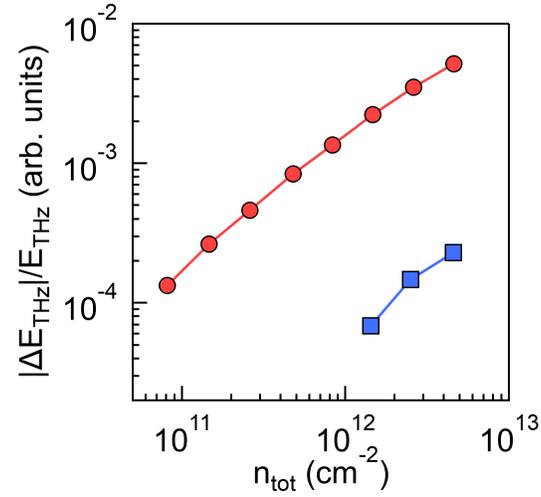

**Supplementary Fig. 7. Excitation density dependence of the THz photoconductivity for MoSe$_2$.** The red circles and blue squares show the results for the CVT and flux-grown MoSe$_2$, respectively. Note that the absolute magnitude of the THz signal is plotted because the signs of THz photoconductivity are different for the CVT and flux samples. These are the peak intensities of the THz TDS scans.

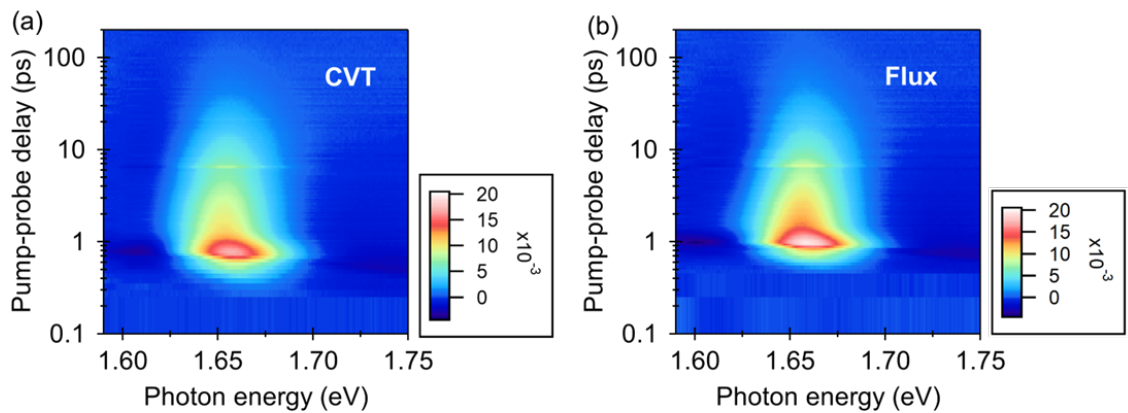

**Supplementary Fig. 8. Transient absorption dynamics at visible frequency.** (a,b) 2D pseudo-color plot of the transient absorption dynamics for the CVT and flux-grown WSe$_2$, respectively. The excitation density was n$_{abs}$ = 5.0 x 10$^{12}$ cm$^{-2}$. The bleaching dynamics monitored at the exciton resonance of 1.66 eV are given in Fig. 4d.

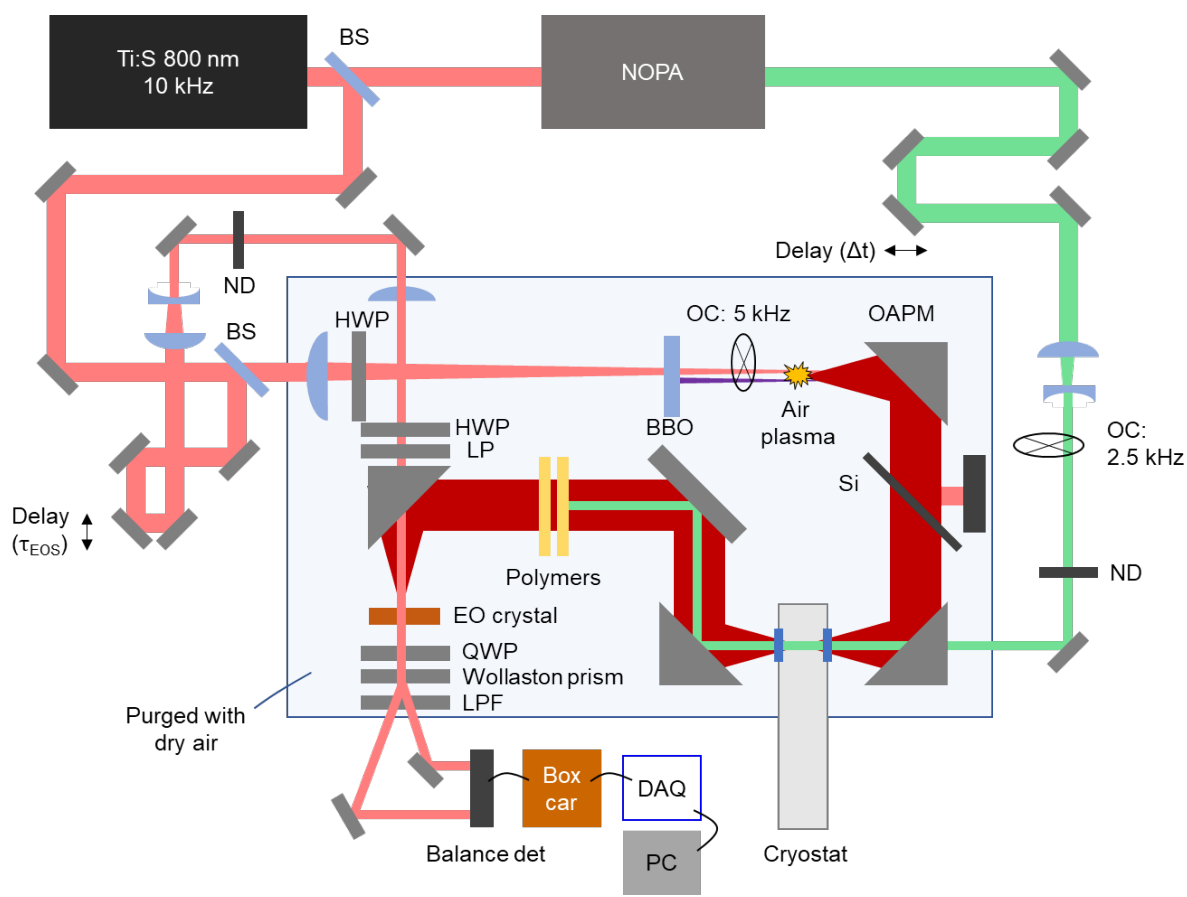

**Supplementary Fig. 9. Schematic for the optical pump-THz probe spectroscopy setup**. ND, neutral density filter; OAPM, off-axis parabolic mirror; HWP (QWP), half (quarter) wave plate; EO, electro-optic; OC, optical chopper; BS, beam splitter; DAQ, data acquisition.

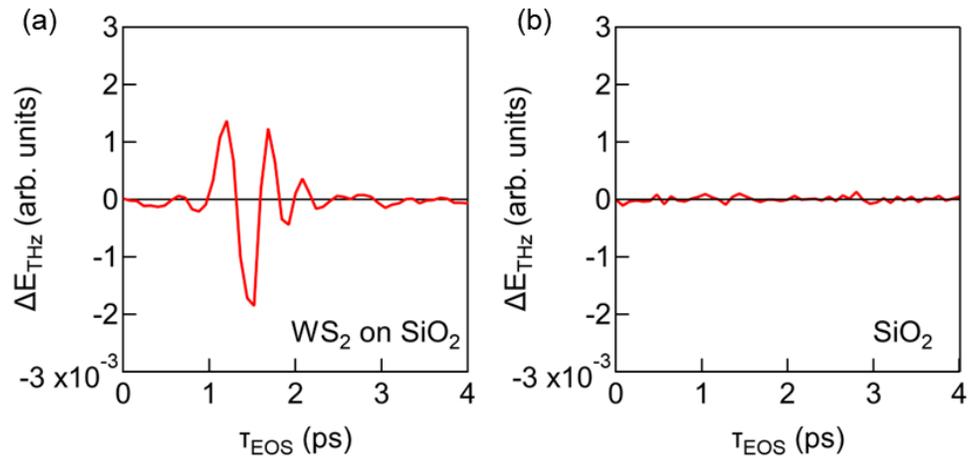

**Supplementary Fig. 10. Negligible signal from the SiO₂ substrate.** (a,b) The photoinduced THz traces for the WS₂ on SiO₂ and the area with only SiO₂, respectively. The excitation photon energy was 2.4 eV, the above-gap excitation condition for WS₂, see Fig. 1b. We confirmed that there is no photoinduced THz signal from the SiO2 substrate by measuring the area without the TMD monolayer.

# References

1. Steinleitner, P. *et al.* Direct Observation of Ultrafast Exciton Formation in a Monolayer of WSe 2. *Nano Lett.* **17**, 1455–1460 (2017).

2. Kormányos, A. *et al.* K.p theory for two-dimensional transition metal dichalcogenide semiconductors. *2D Mater.* **2**, 022001 (2014).

3. Chernikov, A. *et al.* Exciton Binding Energy and Nonhydrogenic Rydberg Series in Monolayer WS2. *Phys. Rev. Lett.* **113**, 076802 (2014).

4. Pedersen, T. G. Exciton Stark shift and electroabsorption in monolayer transition-metal dichalcogenides. *Phys. Rev. B* **94**, 125424 (2016).

5. Kaindl, R. A., Hägele, D., Carnahan, M. A. & Chemla, D. S. Transient terahertz spectroscopy of excitons and unbound carriers in quasi-two-dimensional electron-hole gases. *Phys. Rev. B* **79**, 045320 (2009).

6. Steinhoff, A. *et al.* Exciton fission in monolayer transition metal dichalcogenide semiconductors. *Nat. Commun.* **8**, 1166 (2017).

7. Schmitt-Rink, S., Chemla, D. S. & Miller, D. A. B. Theory of transient excitonic optical nonlinearities in semiconductor quantum-well structures. *Phys. Rev. B* **32**, 6601–6609 (1985).

8. Ceballos, F., Cui, Q., Bellus, M. Z. & Zhao, H. Exciton formation in monolayer transition metal dichalcogenides. *Nanoscale* **8**, 11681–11688 (2016).